\documentclass[12pt]{aastex63}
\usepackage{natbib}
\usepackage{amsmath}
\usepackage{graphicx}
\newcommand{\jb}{\textit{J}}
\newcommand{\hb}{\textit{H}}
\newcommand{\kb}{\textit{K}$_1$}

\begin{document}

\title{Multiband Polarimetric Imaging of HR 4796A with the Gemini Planet Imager}

\author[0000-0001-6364-2834]{Pauline Arriaga}
\affiliation{Department of Physics and Astronomy, University of California, Los Angeles, 430 Portola Plaza, Box 951547, Los Angeles, CA 90095-1547}
\email{parriaga@astro.ucla.edu}

\author[0000-0002-0176-8973]{Michael P. Fitzgerald}
\affiliation{Department of Physics and Astronomy, University of California, Los Angeles, 430 Portola Plaza, Box 951547, Los Angeles, CA 90095-1547}

\author[0000-0002-5092-6464]{Gaspard Duch\^{e}ne}
\affiliation{Astronomy Department, University of California, Berkeley; Berkeley, CA 94720, USA.}
\affiliation{Univ. Grenoble Alpes/CNRS, IPAG, F-38000 Grenoble, France}

\author{Paul Kalas}
\affiliation{Astronomy Department, University of California, Berkeley, CA 94720, USA}
\affiliation{SETI Institute, Carl Sagan Center, 189 Bernardo Ave.,  Mountain View CA 94043, USA}
\affiliation{Institute of Astrophysics, FORTH, GR-71110 Heraklion, Greece}

\author[0000-0001-6205-9233]{Maxwell A. Millar-Blanchaer}
\affiliation{Jet Propulsion Laboratory, California Institute of Technology, 4800 Oak Grove Dr., Pasadena, USA} 
\affiliation{NASA Hubble Fellow}

\author[0000-0002-3191-8151]{Marshall D. Perrin}
\affiliation{Space Telescope Science Institute, 3700 San Martin Drive, Baltimore, MD 21218, USA}

\author[0000-0002-8382-0447]{Christine H. Chen}
\affiliation{Space Telescope Science Institute, 3700 San Martin Drive, Baltimore MD 21218,USA}

\author[0000-0002-9133-3091]{Johan Mazoyer}
\altaffiliation{NFHP Sagan Fellow}
\affiliation{NASA Jet Propulsion Laboratory, California Institute of Technology, Pasadena, CA 91109, USA}
\affiliation{LESIA, Observatoire de Paris, Université PSL, CNRS, Sorbonne Université, Université de Paris, 5 place Jules Janssen, 92195 Meudon, France}


\author[0000-0001-5172-7902]{Mark Ammons}
\affiliation{Lawrence Livermore National Laboratory}

\author[0000-0002-5407-2806]{Vanessa P. Bailey}
\affiliation{Jet Propulsion Laboratory, California Institute of Technology, 4800 Oak Grove Dr., Pasadena, CA 91109, USA 
}

\author{Trafis S. Barman}
\affiliation{Lunar and Planetary Lab, University of Arizona, Tucson, AZ 85721, USA}

\author{Joanna Bulger}
\affiliation{Subaru Telescope, NAOJ, 650 North A’ohoku Place, Hilo, HI 96720, USA}

\author[0000-0001-6305-7272]{Jeffrey K. Chilcote}
\affiliation{Department of Physics, University of Notre Dame, 225 Nieuwland Science Hall, Notre Dame, IN, 46556, USA}

\author{Tara Cotten}
\affiliation{Department of Physics and Astronomy, University of Georgia, Athens, GA 30602, USA}

\author[0000-0002-4918-0247]{Robert J. De Rosa}
\affiliation{European Southern Observatory , Alonso de Cordova 3107, Vitacura, Santiago, Chile
}

\author{Rene Doyon}
\affiliation{Institut de Recherche sur les Exoplanètes, D\'{e}partement de physique, Universit\'{e} de Montréal, Montr\'{e}al, QC H3C 3J7, Canada}

\author[0000-0002-0792-3719]{Thomas M. Esposito}
\affiliation{Astronomy Department, University of California, Berkeley; Berkeley, CA 94720, USA
}

\author[0000-0002-7821-0695]{Katherine B. Follette}
\affiliation{Physics and Astronomy Department, Amherst College, 21 Merrill Science Drive, Amherst, MA 01002, USA}

\author[0000-0003-3978-9195]{Benjamin L. Gerard}
\affiliation{University of Victoria, Department of Physics and Astronomy, 3800 Finnerty Rd, Victoria, BC V8P 5C2, Canada}
\affiliation{National Research Council of Canada Herzberg, 5071 West Saanich Rd, Victoria, BC, V9E 2E7, Canada}

\author{Stephen Goodsell}
\affiliation{Department of Physics, Durham University, Stockton Road, Durham DH1, UK 2. Gemini Observatory, Casilla 603, La Serena, Chile}

\author{James R. Graham}
\affiliation{Astronomy Department, University of California, Berkeley; Berkeley, CA 94720, USA}

\author[0000-0002-7162-8036]{Alexandra Z. Greenbaum}
\affiliation{Department of Astronomy, University of Michigan, Ann Arbor, MI 48109, USA}

\author[0000-0003-3726-5494]{Pascale Hibon}
\affiliation{European Southern Observatory , Alonso de Cordova 3107, Vitacura, Santiago, Chile
}

\author[0000-0001-9994-2142]{Justin Hom}
\affiliation{School of Earth and Space Exploration, Arizona State University, PO Box 871404, Tempe, AZ 85287, USA}

\author{Li-Wei Hung}
\affiliation{Natural Sounds and Night Skies Division, National Park Service, Fort Collins, CO 80525, USA}

\author{Patrick Ingraham}
\affiliation{Large Synoptic Survey Telescope, 950 N Cherry Ave, Tucson AZ, 85719, USA}

\author[0000-0002-9936-6285]{Quinn M. Konopacky}
\affiliation{Center for Astrophysics and Space Sciences, University of California, San Diego, La Jolla, CA 92093, USA}

\author[0000-0003-1212-7538]{Bruce A. Macintosh}
\affiliation{Kavli Institute for Particle Astrophysics and Cosmology, Department of Physics, Stanford University, Stanford, CA, 94305, USA}

\author{J\'{e}r\^{o}me Maire}
\affiliation{Center for Astrophysics and Space Sciences, University of California, San Diego, La Jolla, CA 92093, USA}

\author[0000-0001-7016-7277]{Franck Marchis}
\affiliation{SETI Institute, Carl Sagan Center, 189 Bernardo Av, Suite 200, Mountain View CA 94043, USA}

\author[0000-0002-5251-2943]{Mark S. Marley}
\affiliation{Space Science Division, NASA Ames Research Center, Mail Stop 245-3, Moffett Field CA 94035, USA}

\author[0000-0001-7016-7277]{Christian Marois}
\affiliation{National Research Council of Canada Herzberg, 5071 West Saanich Rd, Victoria, BC, V9E 2E7, Canada}
\affiliation{University of Victoria, Department of Physics and Astronomy, 3800 Finnerty Rd, Victoria, BC V8P 5C2, Canada}

\author[0000-0003-3050-8203]{Stanimir Metchev}
\affiliation{Department of Physics and Astronomy, Centre for Planetary Science and Exploration, The University of Western Ontario, London, ON N6A 3K7, Canada}
\affiliation{Department of Physics and Astronomy, Stony Brook University, Stony Brook, NY 11794-3800, USA}

\author[0000-0001-6975-9056]{Eric L. Nielsen}
\affiliation{Kavli Institute for Particle Astrophysics and Cosmology, Department of Physics, Stanford University, Stanford, CA, 94305, USA}

\author[0000-0001-7130-7681]{Rebecca Oppenheimer}
\affiliation{American Museum of Natural History, Depratment of Astrophysics, Central Park West at 79th Street, New York, NY 10024, USA}

\author{David W. Palmer}
\affiliation{Lawrence Livermore National Laboratory, 7000 East Ave, Livermore, CA, 94550, USA}

\author{Jenny Patience}
\affiliation{School of Earth and Space Exploration, Arizona State University, PO Box 871404, Tempe, AZ 85287, USA}

\author{Lisa A. Poyneer}
\affiliation{Lawrence Livermore National Laboratory, 7000 East Ave, Livermore, CA, 94550, USA}

\author{Laurent Pueyo}
\affiliation{Space Telescope Science Institute, 3700 San Martin Drive, Baltimore, MD 21218, USA}

\author[0000-0002-9246-5467]{Abhijith Rajan}
\affiliation{Space Telescope Science Institute, 3700 San Martin Drive, Baltimore, MD 21218, USA}

\author[0000-0003-0029-0258]{Julien Rameau}
\affiliation{Institut de Recherche sur les Exoplan\'{e}tes, D\'{e}partement de physique, Universit\'{e} de Montréal, Montr\'{e}al, QC H3C 3J7, Canada}

\author[0000-0002-9667-2244]{Fredrik T. Rantakyr\"o}
\affiliation{Gemini Observatory, Casilla 603, La Serena, Chile}

\author[0000-0003-2233-4821]{Jean-Baptiste Ruffio}
\affiliation{Kavli Institute for Particle Astrophysics and Cosmology, Stanford University, Stanford, CA, 94305, USA}

\author[0000-0002-8711-7206]{Dmitry Savransky}
\affiliation{Sibley School of Mechanical and Aerospace Engineering, Cornell University, Ithaca, NY 14853, USA}

\author{Adam C. Schneider}
\affiliation{Physics and Astronomy, University of Georgia, 240 Physics, Athens, GA 30602, USA}

\author{Anand Sivaramakrishnan}
\affiliation{Space Telescope Science Institute, 3700 San Martin Drive, Baltimore, MD 21218, USA}

\author[0000-0002-5815-7372]{Inseok Song}
\affiliation{Department of Physics and Astronomy, University of Georgia, Athens, GA 30602, USA}

\author[0000-0003-2753-2819]{Remi Soummer}
\affiliation{Space Telescope Science Institute, 3700 San Martin Drive, Baltimore, MD 21218, USA}

\author{Sandrine Thomas}
\affiliation{Large Synoptic Survey Telescope, 950 N Cherry Ave, Tucson AZ, 85719, USA}

\author[0000-0003-0774-6502]{Jason J. Wang}
\affiliation{Department of Astronomy, California Institute of Technology, Pasadena, CA 91125, USA}
\affiliation{51 Pegasi b Fellow}

\author[0000-0002-4479-8291]{Kimberly Ward-Duong}
\affiliation{Five College Astronomy Department, Amherst College, Amherst, MA 01002, USA}

\author[0000-0002-9977-8255]{Schuyler G. Wolff}
\affiliation{Leiden Observatory, Leiden University, 2300 RA Leiden, The Netherlands}

\begin{abstract}
HR4796A hosts a well-studied debris disk with a long history due to its high fractional luminosity and favorable inclination lending itself well to both unresolved and resolved observations. We present new \jb- and \kb-band images of the resolved debris disk HR4796A taken in the polarimetric mode of the Gemini Planet Imager (GPI). The polarized intensity features a strongly forward scattered brightness distribution and is undetected at the far side of the disk. The total intensity is detected at all scattering angles and also exhibits a strong forward scattering peak. 
We use a forward modelled geometric disk in order to extract geometric parameters, polarized fraction and total intensity scattering phase functions for these data as well as \hb-band data previously taken by GPI. We find the polarized phase function becomes increasingly more forward scattering as wavelength increases. We fit Mie and distribution of hollow spheres grain (DHS) models to the extracted functions. We find that while it is possible to describe generate a satisfactory model for the total intensity using a DHS model, but not with a Mie model. We find that no single grain population of DHS or Mie grains of arbitrary composition can simultaneously reproduce the polarized fraction and total intensity scattering phase functions, indicating the need for more sophisticated grain models. 
\end{abstract}


\section{INTRODUCTION}\label{sec:introduction}

Since the discovery of the first exoplanets nearly 25 years ago~\citep{mayor95}, the field has developed rapidly in an attempt to answer fundamental questions about the formation and evolution of planetary systems. With the advent of large telescopes with extreme adaptive optics systems, it has become possible to directly image exoplanets \citep{macintosh06,beuzit19}, though recent surveys
have shown that the occurrence rate of planets with a brightness and star separation that can be currently be detected by
direct imaging methods is fairly small \citep{bowler18}. However, the same technology also allows for a different approach to understanding
planetary system architecture and dynamics through the study of the
resolved structure of circumstellar disks. Resolved images of disks show that features such as
sharp radial profiles, warps, clumps and spirals can be caused by unseen planets \citep{nesvold15, quillen06}. Models of observable disk features have led to the discoveries of directly imaged planets around their host stars such as $\beta$-Pictoris b \citep{lagrange10} and Fomalhaut b \citep{kalas05}.

Debris disks are a class of evolved circumstellar disks characterized by low levels of gas and low optical depth. They are mainly composed of
planetesimals and dust, continually replenished by collisions \citep{wyatt08}. This dust allows us to
observe debris disks across many wavelengths as the scattered light can be observed in the optical and near infrared wavelengths
while the emitted thermal light can be observed in the mid-infrared and beyond.

HR4796A is a well-studied debris disk surrounding a 9\,Myr \citep{bell2015} A0V
star, at a distance of $71.91\pm0.70$\,pc from Earth
\citep{van_leeuwen2007,gaia16}. The disk is
exceptionally bright with an infrared excess 
$f = L_{IR} / L_* = 5\times 10^{-3}$ \citep{jura1991} which has made it a popular target for subsequent debris disk studies. Since its discovery the disk has been imaged in many wavelengths from the sub-mm~\citep{sheret04}, the mm~\citep{greaves00}, mid-infrared~\citep{koerner98,lisse17}, near-infrared~\citep{schneider1999,perrin14,milli2017}, and visible~\citep{schneider2009,schneider14,milli19}. These multi-wavelength observations have allowed for extensive modelling of the spectral energy distribution (SED) to understand the dust composition of the disk \citep{li03,rodigas15}. Later studies have resolved a circular disk component at a radius of $\sim$77\,au with a sharp radial profile and a $\sim$1\,au offset from the
star \citep{schneider2009}. The modelling the exact geometry of these features reveals insights on the dynamics of the system ~\citep{wyatt99,wyatt08,nesvold15}.

Resolved imaging additionally provides information about the system through studies of the wavelength-dependent scattering phase functions (SPFs) of the disk-scattered light.
Early total intensity high contrast infrared images by the Gemini Planet Imager (GPI)~\citep{perrin14} had shown hints of an asymmetric brightness distribution with forward scattering peak, which was later fully resolved by the Spectro-Polarimetric High-contrast Exoplanet Research Instrument (SPHERE)~\citep{milli2017}. Though models have not satisfactorily fit the near-IR SPF, such studies have allowed for the elimination of certain gran models such as scattering by submicron Mie particles.

High-contrast imaging instruments have enabled the studies of the polarized intensity of the disk. Polarized images have the advantage of not requiring PSF subtraction of the randomly polarized star's light. \cite{hinkley09} presented the first near-infrared detection of the disk in polarized intensity, finding robust detections at the ansae. Later images high-contrast images taken by GPI \citep{perrin14} fully resolved the disk in polarized intensity.  The images showed a highly asymmetric polarized intensity scattering phase function (SPF), with the disk intensity strongly peaking at the smallest scattering angle and undetected at the largest scattering angles. Recently VLT/SPHERE has imaged the polarized intensity in optical light, similarly showing an asymmetric polarized SPF. The polarized SPF in conjunction with the total intensity SPF allows for tighter constraints on the properties and composition of the scattering dust grains.


In this study, we present new \jb- and \kb-band total and polarized intensity images. We perform modelling on these images as well as the \hb-band polarized intensity image presented in \citep{perrin14}. We aim to expand our knowledge of the polarized and total
intensity phase functions in near-IR and by proxy study the properties of the scattering grains in this system. In
Section~\ref{sec:datared} we describe the observations and the data reduction techniques. In in Section~\ref{sec:geo_modelling} we then construct models parameterized only by geometric parameters remaining agnostic to any underlying physical mechanisms driving the grain orbits. 
Having extracted the scattering phase function and polarized phase function, we then fit physical Mie and distribution of hollow spheres (DHS) grain models to our scattering phase function described in ~\ref{sec:phasefunction}. 

\section{DATA ACQUISITION \& PROCESSING}\label{sec:datared}

\subsection{HR4796A Observations}\label{sec:data_obs}
Gemini Planet Imager (GPI) is a high-contrast imaging instrument built for the Gemini Observatory and has been operating at the Gemini South Telescope\citep{macintosh06}. GPI has an integral field polarimetry mode as well as a integral field spectrograph mode. In the polarimeter mode, light is sampled in the pupil plane by a lenslet array followed by a polarizing beam splitter. The raw image consists of an array of spatial resolution elements or spaxels, with pairs of spots of orthogonal linear polarizations. The light is modulated by a half-wave plate (HWP) between each exposure such that $I$, $Q$, and $U$ images can be constructed from a sequence of exposures.

We observed HR4796A with GPI on March 22, 2014.  A summary of the observations are listed in Table~\ref{tab:obs}. Thirty-five 60-second exposures
were taken in \jb-band ($\lambda_c = 1.24\micron$) polarimetry mode with 65$^\circ$ of field
rotation followed by thirty-eight 60-second exposures in \kb-band ($\lambda_c = 2.05\micron$) with
43.8$^\circ$ of field rotation under good seeing conditions. The
half-wave plate was rotated between position angles 0$^\circ$,
22$^\circ$, 45$^\circ$, 68$^\circ$ throughout each sequence. We additionally used \hb-band ($\lambda_c = 1.65\micron$) polarimetry mode data whose acquisition and reduction is described in \cite{perrin14}.

\begin{table} 
\begin{center}
\begin{tabular}{|c|l|c|c|c|c|c|c|}
\hline
Target & UT Date & Filter & Obs. Mode & No. Exps. & Field Rot. ($^\circ$) & Airmass & Seeing \\
\hline
HR4796A & 2016 Mar 23 & J\_coron & spec & 59 & 48.8 & 1.02 - 1.03 & 0.4 - 0.7\\
\hline
HR4796A & 2016 Mar 18 & H\_coron & spec & 37 & 52.7 & 1.01 - 1.02 & 0.5 \\
\hline
HR4796A & 2015 Apr 3 & K1\_coron & spec & 46 & 78.5 & 1.01 - 1.02 & 0.3 \\
\hline
HR4796A & 2014 Apr 22 & J\_coron & pol & 35 & 65 & 1.03 & 0.3\\ 
\hline
HR4796A & 2014 Apr 22 & \kb\_coron & pol & 38 & 43.8 &  1.02 & 0.7\\
\hline
HR4796A & 2013 Dec 12 & H\_coron & pol & 11 & 2.1 & 1.3 & 0.2 \\
\hline
\end{tabular}
\end{center}
\caption{Observations\label{tab:obs}}
\end{table} 

\subsection{Data Reduction}\label{sec:data_red}
The raw data were reduced using the GPI Data Reduction Pipeline
\citep{perrin14}. The raw images were dark subtracted, flexure
corrected, destriped, and corrected for bad pixels. The orthogonal
polarization spots were then extracted from each raw image to form a
polarization cubes, each with two frames of orthogonal
polarization. The cubes were then divided by a polarized flat field. Bad pixels were identified and replaced via interpolation. The star's
position was measured using the position of 
reference satellite spots diffracted from starlight behind the coronagraph~\citep{wang2014}.

\subsection{Polarized and Angular Differential Imaging}\label{sec:pdi_adi}
Data taken in GPI's polarimetry mode are particularly suited for both polarized and angular differential imaging, both of which suppress the starlight and improve the contrast by orders of magnitude. For polarized differential imaging (PDI), we subtracted the two frames of orthogonal polarization for each datacube, removing the majority of the starlight which has a randomly oriented polarization. Stokes' cubes were then constructed from the resultant frames using a singular value decomposition method \citep{perrin15} to recover Stokes parameters from the data and HWP-modulated time-variable measurement matrices. The mean stellar polarization was corrected for by first measuring the average polarized intensity (Stokes parameter $I$) inside of the focal plane mask. The total intensity image scaled by the measured intensity was then subtracted from the linearly polarized intensity image. The final image that was fit to in subsequent modelling described in Section~\ref{sec:geo_modelling} was a radial polarization image, a combination of the Q and U images that gives the polarization in the radial direction. For an optically thin single scattering disk, all of the signal is expected to lie in this radial polarization.

Another advantage of this polarization data is that the sum of the linear polarization states can be combined and processed using an angular differential imaging \citep[ADI]{marois06} algorithm, to produce a PSF subtracted total intensity image. For each data cube, we combined the two linear polarization states to form a series of total intensity images to correspond to each polarization image. We then used a
Python implementation of Karhunen-Loeve Image Projection 
\citep[KLIP]{soummer12}, PyKLIP \citep{wang2015}, to perform this angular
differential imaging. We
optimized the size and number of subtraction regions and the number of
basis vectors subtracted to minimize PSF self-subtraction of the disk by making measurements of the signal to noise at various points along the disk as a function of KLIP parameters. Our measurements indicated the optimal parameters were one basis vector and one subtraction. 

\subsection{Spectral Mode Observations}\label{sec:psf_obs}
Our forward model as described in Section~\ref{sec:datared} requires a convolution of our model with a point spread function (PSF).
The PSF for GPI is challenging to model due to its complex
structure and variability \citep{wang2014}. As such, rather than use a Gaussian or Airy function, we used a PSF that extracted from the four
satellite spots dispersed in each image of
HR4796A. Since polarimetric frames are broadband and are therefore have overlapping satellite spots in a single frame, we extracted the PSF structure
from satellite spots of observations taken in GPI's integral field spectrograph (IFS) mode. We elected to use the HR4796A satellite spots even though the field is noiseir than that of observations of other stars because of the dependence of the PSF geometry on the stellar spectrum. The stellar spectra would affect the relative weights of the
extracted satellite spots at different wavelengths. We used thirty-six
60-second \hb-band frames taken on March 18, 2016, fifty-eight 30-second
\jb-band frames taken on March 23, 2016, and thirty-six 60-second \kb-band frames
taken on March 13, 2016.

\subsection{Convolution PSF Construction}\label{sec:psf_const}
These data were also reduced using the GPI Data Reduction Pipeline
\citep{perrin14}. The raw images were dark subtracted, flexure
corrected, destriped, corrected for bad pixels. The spectra for each spaxel were then
extracted to form 3-D data cubes. The data cubes were then
further corrected for bad pixels and distortion.

In order to estimate the PSF, we first summed the spectral mode images along the wavelength axis and the polarimetric frames on the polarized axis, both giving estimates of the total intensity across the bandpass. We took the median image of these flattened spectral mode images and the median of the polarimetric mode images. We high-pass filtered both median images with a box size of 6 spaxels, in order to remove
large-scale structure from the main image of the star behind the occulting PSF. This box size was chosen to optimize the uniformity of the background structure surrounding the star. Each spectral channel was linearly combined with a weight to find the least-square difference with the polarimetric satellite spot. The need for different weighting parameters stems from the difference in throughput between spectral and polarimetric mode. We registered the spectral satellite spots of each wavelength, multiplied each one by the weights we had fitted for and summed them along the wavelength axis to get an image of the PSF. Though the PSF is highly asymmetric with lobes at four locations around the core, for our model convolution, we azimuthally averaged this PSF since each image of the disk is derotated for sky
rotation making the final PSF a combination of rotated PSFs from individual exposures.

\section{PRESCRIPTIVE MODELLING}\label{sec:geo_modelling}

\subsection{Model Description}\label{sec:mod_descrip}
In order to extract the geometric parameters and brightness function of the disk, we fit a geometric model to the data. By fitting a model purely generated from an arbitrary description of phase and geometric parameters, we remain agnostic to any assumptions about the physical forces on the dust grains, the orbital grain distribution, or the properties of the dust grains. In this procedure, we additionally use KLIP forward modelling~\citep{pueyo16} to account for self-subtraction of the disk brightness in total intensity. 

We selected our preferred prescription for the disk by minimizing the number of parameters needed to achieve comparable $\chi^2$ values. We found that modeling the disk as an ellipse as opposed to an offset circle added extra parameters that did not improve the $\chi^2$ sufficiently to warrant the more complicated disk. We therefore modeled the ring as a series of nested circles offset from the star. We fit for $\Omega$, the position angle of the major axis, and $i$, the
inclination of the nested circles. 

We constructed the model by mapping each pixel location $(x,
y)$ in the sky plane to a radius $r(x, y) = \sqrt{x_{\text{disk}}(x,y)^2 +
  y_{\text{disk}}(x,y)^2}$ in the disk plane and a $\theta(x, y) =
\tan^{-1}(x/y)$ in the sky plane. The intensity of each pixel is then   
\begin{equation}
    I(x, y) = \frac{B_r(r) B_\theta(\theta)}{r_*^2},
\end{equation}
where $B_r$ is a radial profile, $B_\theta$ is the azimuthal intensity profile, and $r_*$ is a unitless factor that scales with the distance between the location (x, y) and the star. $B_\theta$ is a periodic spline interpolation with varying
numbers of knots, with the intensity at every knot as a free
parameter. Transforming $B_\theta$ to $B_\phi$ where $\phi$ is the scattering angle gives the SPF of each model disk. As all of the resulting phase curves are normalized in analysis in Section~\ref{sec:phasefunction}, the units on I(x, y) are arbitrary.

\begin{figure}
\centering%
\includegraphics[height=.4\textheight]{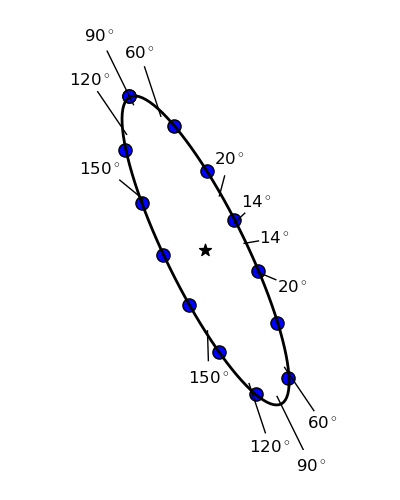}%
\caption{The blue points indicate the locations of the spline points of the intensity of the disk around the disk. The number of spline points was determined to be the minimum number of points to recover a known SPF to a 1\% level. The marked angles are the scattering angles assuming that the west side is closer to the observer. The extracted polarized phase function was cut off at a scattering angle of $120^\circ$ where the disk signal falls below the noise level. The total intensity data was truncated at angles less than $20^\circ$ and greater than 150$^\circ$.}
\label{fig:cartoon}%
\end{figure}

The knots are evenly spaced in the sky plane along the disk as shown in the blue points in Figure~\ref{fig:cartoon}. By spacing the points with a larger separation than the scale of the PSF, we minimized the effects of spatially correlated noise residuals. To find the optimal number of spline points, we used the dust modelling package MCFOST~\citep{pinte06} to generate a model with similar geometry to HR4796A and a known two-component Henyey-Greenstein phase function with the dust modelling package MCFOST. We then injected this model into a separate polarization dataset with no disk detection. We used our modelling procedure to recover this artificial disk. Transforming the recovered $B_\theta$ to $B_\phi$ gave a curve that could be directly compared to a theoretical $B_\phi$ scattering phase function. Using a minimum number of 13 knots we were able to recover the scattering phase function to the 1\% level at all observable scattering angles. In this test injection, the fit was invariant to the location of the spline points as long as there were a sufficient amount of points to fully describe the shape. However, this test assumes that the disk has a smooth phase function at all points, which, as will be shown in Section~\ref{sec:phasefunction} may not be the case. It also assumes that there is no large-scale noise structure that the model would fit to, which is fortunately the case for most of the images we have modelled. 
Though the locations of the spline points are not densely sampled in $\phi$ vs $B_\phi$ space, as long as the intrinsic SPF is smooth, the spline will recover its shape at all accessible scattering angles. This is in contrast to extractions of an SPF that use aperture photometry to sample to the brightness $B_\theta$ which can only be described by brightnesses at the discrete locations of the apertures and suffer from self-subtraction bias. 

The radial profile $B_r$ is a broken power law:


\begin{equation}
B_r(r) = \begin{cases}
     \left(\frac{r}{r_0}\right)^{-\gamma_{\text{in}}} & \text{if } r_{\text{in}} < r < r_0 \\
     
     \left(\frac{r}{r_0}\right)^{-\gamma_{\text{out}}}& \text{if } r_0< r < r_{\text{out}} \\
     0 & \text{otherwise}
    \end{cases}
\end{equation}
where $r_0$, the central radius in milli-arcseconds, $\gamma_{\text{in}}$, the inner radial profile, and $\gamma_{\text{out}}$, the outer radial profile, as
free parameters. The radial profile was found to be very sharp to the point that the $\gamma$ factors were degenerate with the inner and outer radii. To reduce the number of parameters and avoid unbounded parameters $r_{in}$ and $r_{out}$ were fixed at 70 and 100au respectively. These radii were selected by deprojecting the disk and finding the radii where the SNR of the disk falls below 10\%.


\subsection{Fitting Procedure}\label{sec:geo_fitting}
We then used our model disks to fit to the \jb- and \kb-band total and polarized intensity images and the \hb-band polarized intensity image. The \hb-band total intensity did not have enough field rotation to reliably be forward modelled. We created model images with the above parameters, which we then convolved with our derived convolution PSF. In order to simulate the effects on the final data product due to the KLIP PSF subtraction, we developed the DiskFM module for PyKLIP~\citep{wang2014}, specifically for forward modeling extended objects based on the mathematical framework presented in \cite{pueyo16}. Due to this extra step of modeling, we fixed the geometrical parameters of the total intensity disks to those found from their polarized intensity counterparts and only fit the scattering phase spline function. This is a natural choice as the total and polarized intensity images are generated from the same raw data images.

We fit each disk's geometric parameters independently from band-to-band to account for various physical and non-physical effects. The position angle of the line of nodes ($\Omega$) and inclination ($i$) could differ between bands as there is some uncertainty in the rotation of the instrument relative to north. The radial profile parameters were fitted separately to reflect possible changes between the distributions of differently sized dust grains due to differing effects of radiation pressure and gravitational forces.


We independently fit the by using a linear least squares algorithm to optimize the $\chi^2$ using the uncertainty maps. We then used the resultant parameters to seed a fit using an ensemble Markov-chain Monte Carlo (MCMC) using the \textit{emcee} package \citep{emcee}. The final geometric parameters are shown in Table~\ref{tab:geo_parameters} with their error bars derived from the distributions of the final walkers. After fitting for the geometrical parameters, we fixed all of the geometrical parameters for each model disk and fit only the spline parameters. 

\begin{table} 

\begin{center} 
\begin{tabular}{|c|c|c|c|c|c|c|}
\hline
Parameter & \kb~Pol & \hb~Pol & \jb~Pol & Milli 2017 & Schneider 2018 & Units \\
\hline
PA & $27.12\pm0.12$ & $27.14\pm0.12$ & $27.59\pm0.12$ & $27.1\pm 0.7$ & $26.37\pm 0.22$ &  $\degr$ \\
\hline
$i$ & $76.53\pm0.08$ & $76.57\pm0.15$ & $76.91\pm0.12$ & $76.45 \pm 0.7$ & $75.92\pm0.14$ & $\degr$ \\
\hline
$\gamma_{\rm out}$ & $-15.87\pm0.19$ & $-14.13\pm0.21$ & $-13.58\pm0.12$ &  &  & \\
\hline
$\gamma_{\rm in}$ & $42.5\pm0.79$ & $54.73\pm0.66$ & $37.0\pm0.30$ &  &  & \\
\hline
$\omega$ & $-70.37\pm0.38$ & $-72.9\pm.33$ & $-62.6\pm0.18$& & & $\degr$\\
\hline
offset & $52.01\pm 0.49$ & $62.370\pm 12.24$ & $17.04 \pm 13.31$ & & & mas \\
\hline
$r_0$ & $1062\pm3.19$ & $1053\pm3.65$ & $1064\pm3.45$ & $1064\pm6$ & $1059\pm4.6$ & mas \\
\hline

\end{tabular}
\caption{Best fit geometric parameters of the model fits to the polarized intensity images, with 3$\sigma$ errors. We chose to fit the PA and inclination separately for each disk to account for uncertainty in rotation of the instrument relative to north due to the instrument in different cycles as well as differences due to flexure. Additionally, we fit the radial profile parameters to account for possible differences in the structure of the disk of different grain sizes. The fifth column lists the parameters found, averaged over H, H2 and H3 bands, by ~\cite{milli2017} and the sixth lists the average parameters found by ~\cite{schneider14} with the F25ND3 filter. 
The radius for ~\cite{milli2017} is the average distance from the star of points along an ellipse with a semi-major axis $a = 1.066"$ and an ellipticity $e = 0.07$. \label{tab:geo_parameters} }
\end{center} 

\end{table}


\section{DISK GEOMETRY RESULTS}\label{sec:geo_results}
\subsection{Geometric Parameters}\label{sec:geo_params}
\begin{figure}
    \centering
    \includegraphics[height=.8\textheight]{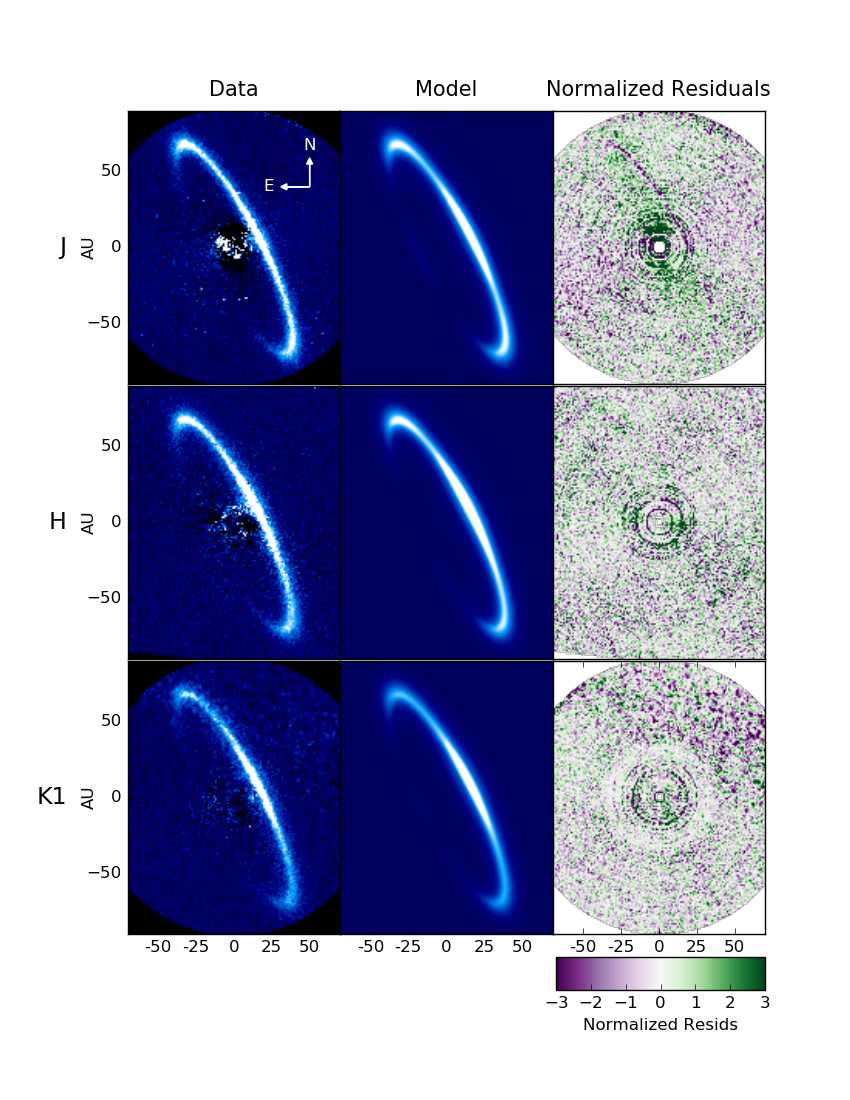}
    \caption{The left column shows the data, while the middle column shows the best fit model for the polarized intensity image. The right column shows the model subtracted from the data divided by the noise map. While most of the normalized residuals indicate per-pixel $\chi^2$ under 1, the \jb-band image shows some structure.} 
    \label{fig:poldata}
\end{figure}

The data and best-fit models for the polarized intensity data are shown in the left and middle columns of Figure~\ref{fig:poldata}. The third column shows the difference between the final and data images divided by our noise map. The residuals for the \hb- and \kb- band model are consistent with the data estimated data uncertainties. Some residual structure may be seen in the \jb-band image north-east ansa, which we will later discussion Section~\ref{sec:rad_prof}. 

\begin{figure}
    \centering
    \includegraphics[width=\textwidth]{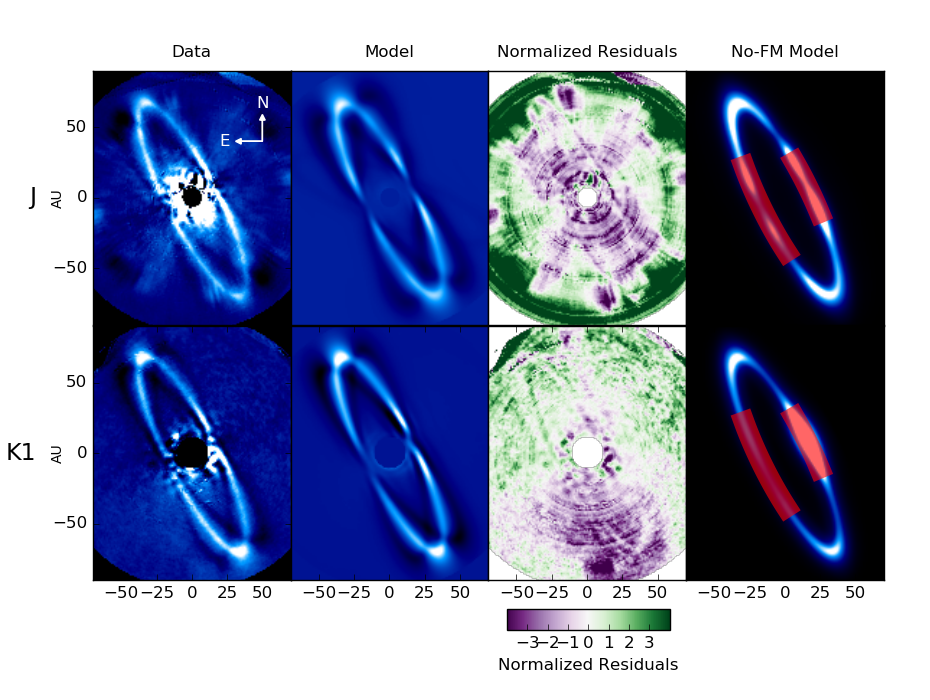}
    \caption{Left column: KLIP PSF subtracted \jb- and \kb-band data images. Middle left column: Best fit forward model. Middle right column: Forward model subtracted from data divided by the noise map. Right column: convolved model before forward modelling. The shaded regions indicate areas we have omitted in our phase curve fits. Upper row: \kb-band. Lower row: \jb-band. We chose not to fit to the \hb-band total intensity due to the small amount of field rotation.}
    \label{fig:totdata}
\end{figure}

\begin{figure}
    \centering
    \includegraphics[width=\textwidth]{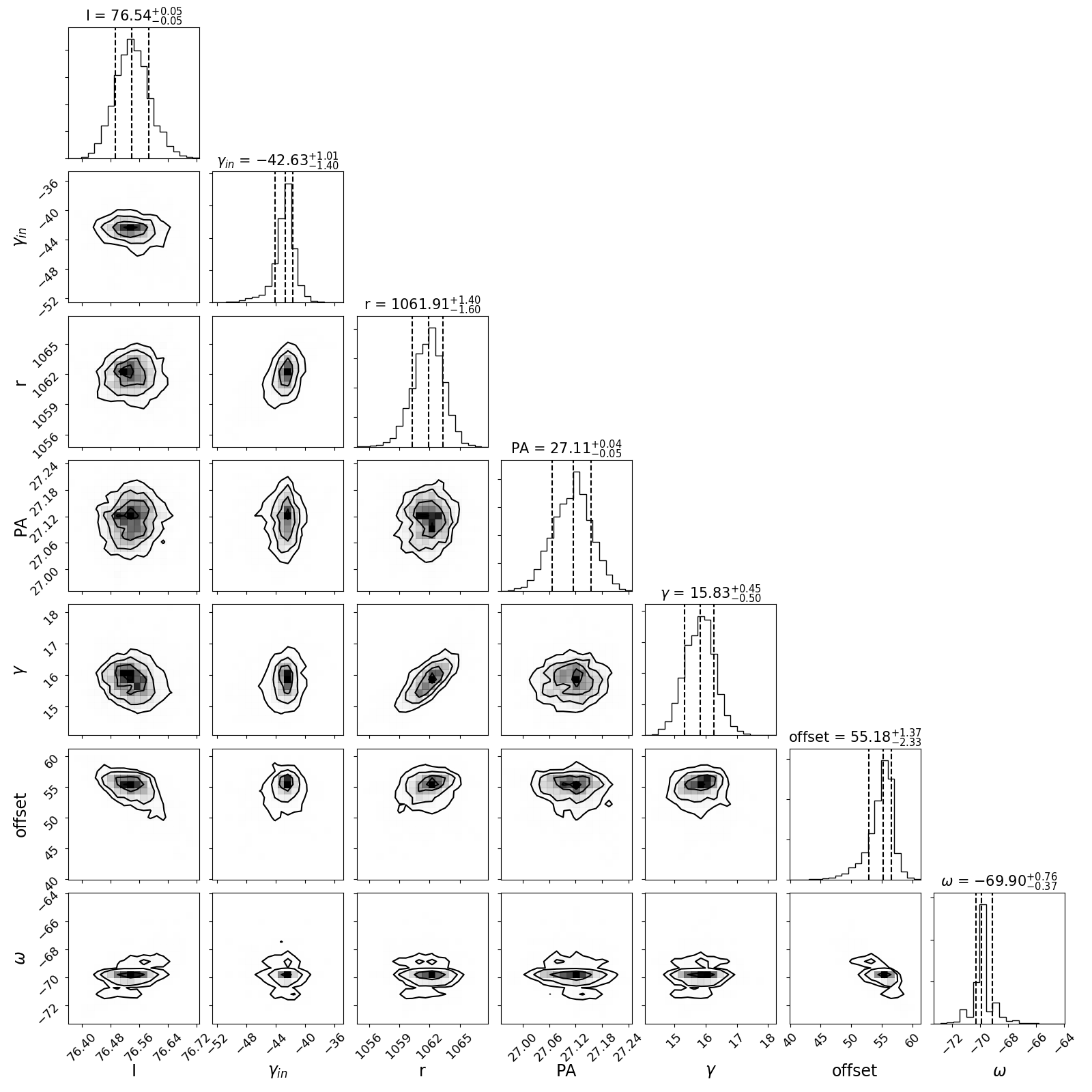}
    \caption{Probability density distribution for the \kb-Pol parameters shown in Table~\ref{tab:geo_parameters}. \hb-band and \jb-band polarized intensity show similar structure, though with wider distributions due to lower signal-to-noise.}
    \label{fig:corner}
\end{figure}

Fits to the total intensity data are shown in Figure~\ref{fig:totdata}. 
The northwest portion of the model in the J-band image overfitted to speckle noise, most evident in the image of the model without forward modeling which shows a likely unphysical dip in intensity. Because of this, we have decided to omit the $\jb$-band polarized fraction and total intensity curves from the phase curve analysis in Section~\ref{sec:phasefunction}. 

The final distributions of the MCMC walkers for the \kb-band fit are shown in Figure~\ref{fig:corner} and the best fit parameters with $3\sigma$ variance for all bands are listed in Table~\ref{tab:geo_parameters}. It is evident from both the walker distributions in Figure~\ref{fig:corner} that the variance of the final parameters are unrealistically small, most likely due to some model mismatch. In the final values for the PA and $\omega$ (the direction of the offset), listed in Table~\ref{tab:geo_parameters}, we have included the variance of the image from true north of $-1^\circ \pm .001^\circ$ found by \cite{konopacky2014}. Calculations of the radius in milli-arcseconds shown in the last line of the table have included the error in assumed plate scale of $14.14\pm10^{-5}$ milli-arcseconds.

Table~\ref{tab:geo_parameters} also shows comparisons to parameters found by \cite{milli2017} in \hb band and \cite{schneider14} in the F25ND3 filter. \cite{milli2017} had found their geometric parameters by fitting radial profiles to cuts of the image and fitting ellipses through the maximal radial values of every profile. To compare with our circular model, Table~\ref{tab:geo_parameters} shows the average distance of every point along the ellipse to the star with their best fit parameters of a semi-major axis $a = 1066mas$ and ellipticity $e = 0.07$. As there is strong residual structure in the \jb-band image which is likely driving the fit parameters, it is most useful to compare parameters between \kb- and \hb-band parameters. The geometry parameters of the position angle and inclination in these bands are consistent not only with each other but also with ~\cite{milli2017}. The radii found in ~\cite{milli2017} are consistent with our derived \kb- and \jb-band models, but not in the \hb-band model. This may be due to biasing of fit by noise close to the focal plane mask in \hb band. Overall, the most consistent and reliable geometric measurements come from \kb band.  


\subsection{Radial Profile}\label{sec:rad_prof}

\begin{figure}
    \centering
    \gridline{
        \fig{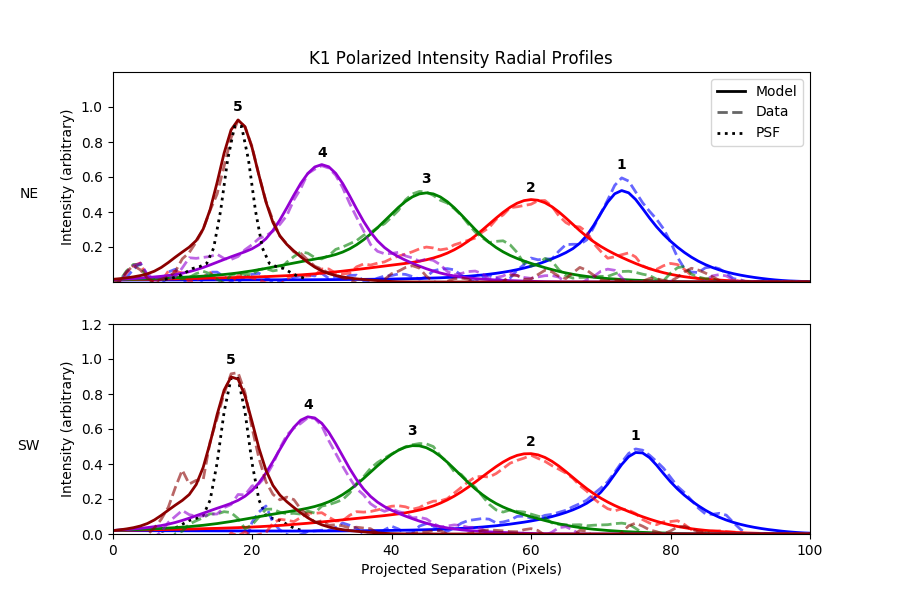}{0.49\textwidth}{(a) \kb-band} 
        \fig{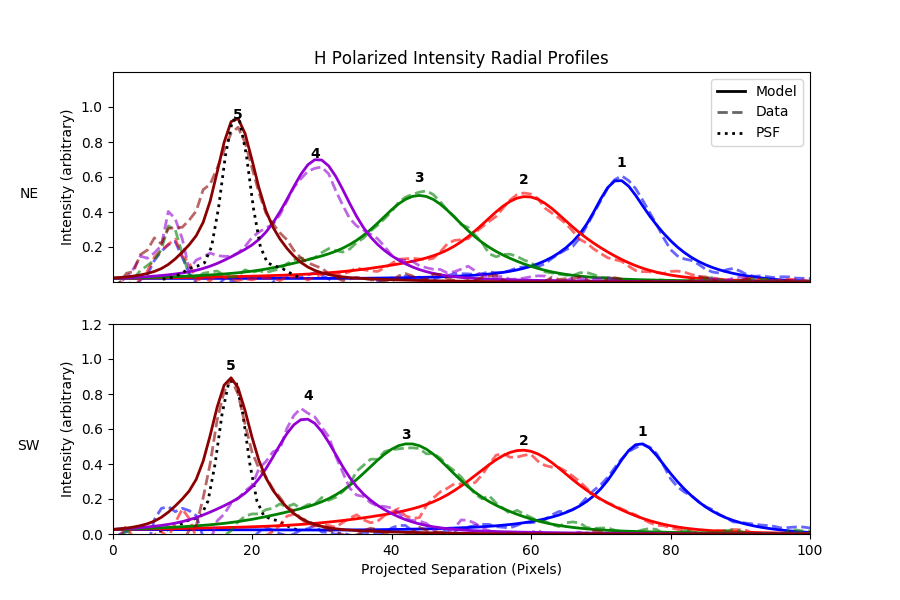}{0.49\textwidth}{(b) \hb-band }
    }
    \gridline{
        \fig{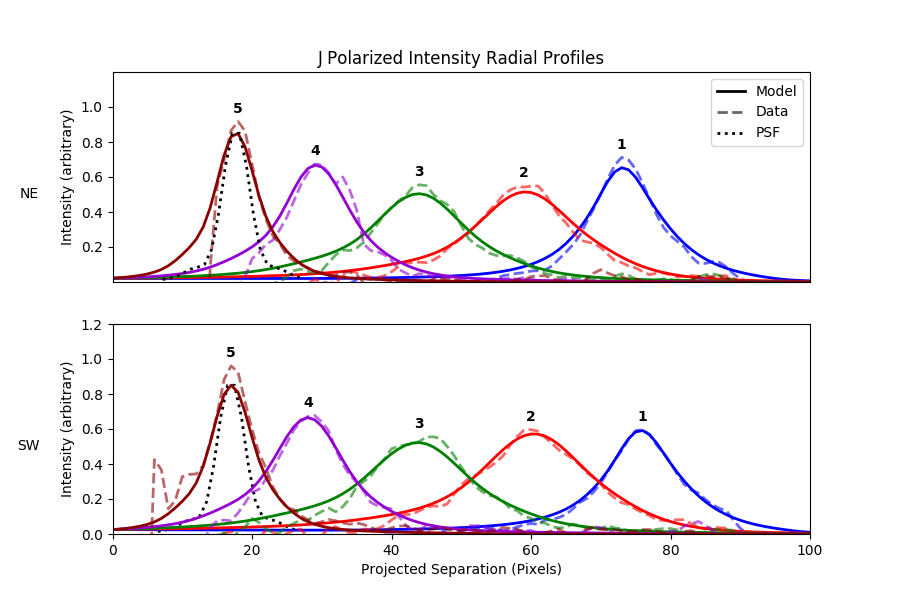}{0.49\textwidth}{(c) \jb-band }
        \fig{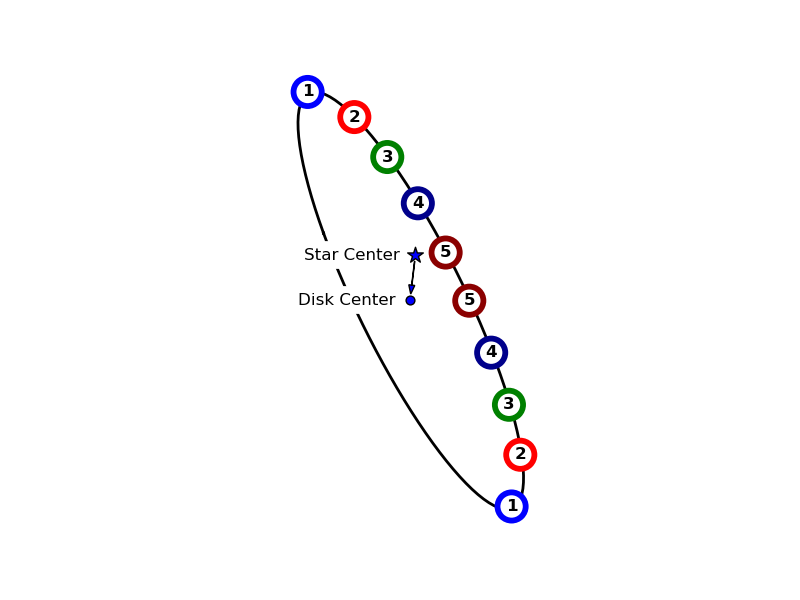}{0.49\textwidth}{(d) Projected radial locations of the radial cuts taken}
    }
    \caption{Radial cuts of the data and model. As a comparison, the cross-section of the PSF is shown in the dotted black line in each plot. Each of the data and model images were rotated such that the x-axis aligned with one of the radial locations marked in (d), which additionally shows the direction of the star center from the disk center with an exaggerated distance. A horizontal cut 4 pixels deep along the x axis were summed along the vertical axis. The solid lines show the cuts of the model images while the dashed lines are the cuts of the data images at each of the radial locations. }
    \label{fig:radial_cuts}
\end{figure}
The inner radial profile exponents $\gamma_{in}$ are large, indicating an unresolved inner edge. The error bars are unrealistically small due to non-uniformity in the radial profile, with one ansae forcing a steeper radial profile and the other forcing a broader radial profile. This effect is most visible in the \jb-band residual image in the northern portion of the disk in the upper left panel of Figure~\ref{fig:poldata}. There are residuals between the data and the model exactly at the midplane, and a visual inspection of the \jb-band data and model show that this is likely due to the data's radial profile being sharper in that region. A more direct representation of the radial profile can be seen in Figure~\ref{fig:radial_cuts}, in which we plotted the intensity radially along the cuts in the directions shown in Figure~\ref{fig:radial_cuts}(d). The \kb-band radial profiles in Figure~\ref{fig:radial_cuts}(a) show little systematic deviation between the model and data. In the \hb-band radial profiles, the radial profiles near the ansae are well fit, though there is evidently noise at small scattering angles near the focal plane mask. This is a likely explanation for the small radius in the \hb-band fit. In \jb-band shown in Figure~\ref{fig:radial_cuts}(c), the southwest radial fits are good, but the peaks of the model cuts are systematically lower than those of the data in the NE region. The radial profile fit is forced by the inner and outer sides of the profiles, lowering the peak. As the model's radial profile is uniform about the disk, this indicates that the radial profile in the north-east half of the disk is sharper than in the south-east half. Such an effect would be seen most evidently in \jb-band as it has the smallest PSF and highest resolution. Qualitatively, the narrowing of the disk in the north-west side is consistent with the \cite{olofsson2019} measurements of polarized SPHERE/ZIMPOL data. We refer the reader to \cite{olofsson2019} for an in-depth discussion of the physical mechanisms possibly causing this effect.

\section{Phase Function\label{sec:phasefunction}}
\subsection{Phase Curve Extraction Results}
\begin{figure}
    \centering
    \includegraphics[width=0.8\textwidth]{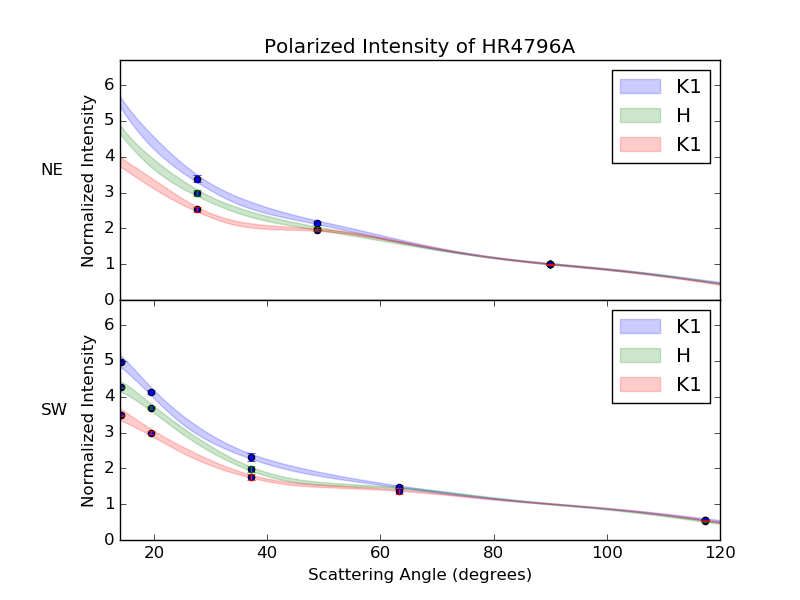}
    \caption{Modelled polarized intensity phase curves as a function of scattering angle. The data points show the locations of the fitted spline points. The 3-$\sigma$ data point error bars are overlaid, but are smaller than or equal to the size of the points. 
The shaded regions represent the 3-$\sigma$ range of the phase curve, derived from the scatter of the splines generated from each MCMC walker's spline point values. 
 The curves are truncated at 120$^\circ$ where the signal is dominated by the noise.
 }
    \label{fig:data_phase}
\end{figure}

 The polarized intensity curves are shown in Figure~\ref{fig:data_phase}, normalized at a scattering angle of $90^{\circ}$. The scattering phase functions are strongly forward scattering with both the polarized and total intensity phase curves peaking at the smallest scattering angles. The NE and SW curves in \hb- and \kb-bands are symmetric, while the NE ansa of the \jb-band image has a bump at a scattering angle of $55^\circ$ due to the residual structure seen in the images at this scattering angle. While the phase curves have similar behavior from $70^\circ - 120^\circ$, the height of the peaks vary with wavelength. The \kb-band phase curve ($\lambda_c = 2.05\micron$) evidently has a sharper forward scattering peak than the \hb-band's ($\lambda_c = 1.65\micron$) which is sharper than the \jb-band's ($\lambda_c = 1.12\micron$). Polarized intensity phase curves taken by ZIMPOL at $\lambda_c = 0.74\micron$ shows that this effect extends to smaller wavelengths, with the phase curve similarly decreasing from $80^\circ - 120^\circ$ but plateauing from $13^\circ - 80^\circ$ ~\citep{milli19}. The source of this chromaticity is unknown as it is plausible for the effect to be caused by a different spatial distribution of multiple grain populations or chromatic effects of a single dust population. Since we are only analyzing the polarized intensity and not the polarized fraction of the \jb- and \hb-band datasets, consistency cannot be checked for chromatic effects with modelling.

\begin{figure}
    \centering
    \includegraphics[width=0.8\textwidth]{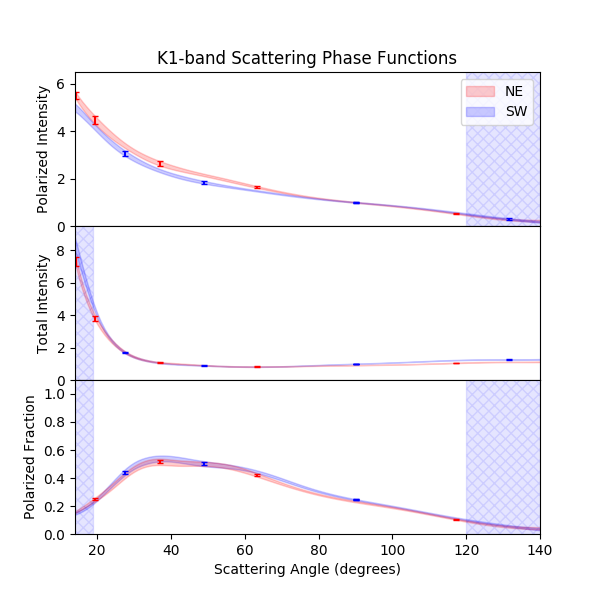}
    \caption{Upper panel: \kb-band polarized intensity phase curves. Lower panel: \kb-band total intensity phase curves. These curves are normalized at 13$^\circ$ and the error bars are derived from the scatter of the MCMC walkers' splines. The shaded portions at $20^\circ$ and $140^\circ$ were excluded from analysis where the signal to noise is low due to attenuation by the PSF subtraction. The data points indicate the locations of the curve spline points. Lower panel: \kb-band polarized fraction. The polarized fraction is derived by dividing the unnormalized polarized intensity phase curve from the unnormalized total intensity phase curve. The south-west curves are plotted in blue, while the north-east curves are plotted in red.}
    \label{fig:ktot_phase}
\end{figure}
 The \kb-band total intensity and polarized fraction curves are shown in the middle and bottom panels, respectively in Figure~\ref{fig:ktot_phase}. The total intensity curves were normalized at 1 at $90^\circ$, while the polarized fraction is unitless. Though it is challenging to measure the polarized and total intensities in physical units, the unnormalized curves can be divided to calculate the polarized fraction. Consistent with phase curves in a similar band in ~\cite{milli2017} the total intensity phase curve exhibits a forward scattering peak and a flat distribution rising at scattering angles larger than $70^\circ$. The polarized fraction curve peaks at $\sim 40^\circ$ at 50\%, consistent with the lower limit found by \cite{hinkley09} of 44\% and the peak polarization found by ~\cite{perrin14} of 50\% at a scattering angle of $50^\circ$.

\subsection{Dust Grain Modelling}\label{sec:grain_model}
We used the MCFOST package \citep{mcfost} to generate theoretical Mie and distribution of hollow spheres (DHS) phase functions \citep{min2005} to fit to our measured phase functions. We modelled to our highest fidelity curves, the south-west \kb-band total intensity and the polarized fraction. We used MCFOST to compute total intensity and polarized fraction phase curves using a given set of parameters at the central wavelength of the \kb band filter. Since the change in grain properties over the \kb band filter is small for most materials, integrating over the whole band did not significantly affect the morphology of the curve. For the total intensity curves, we compared the data to a scaled model where we found the scaling factor by taking the ratio of the model and data curves at every scattering angle and taking the median of those ratios. 

Using the scaled total intensity curve and the polarized fraction then computed reduced $\chi_\nu^2$ values for each curve. As the profiles were generated from a previous fitting procedure we expect the errors to be correlated, but this was ignored in this $\chi^2$ calculation. In the total intensity fit, we excluded regions at scattering angles smaller than $20^\circ$ and larger than 140$^\circ$ as the data are unreliable close to the focal plane mask, shown in Figure~\ref{fig:ktot_phase}. We truncated the polarized fraction curve past 120$^\circ$ as the signal of the polarized intensity is undetected. The locations of these cut-off scattering angles with respect to the disk are shown in Figure~\ref{fig:cartoon}.

We ran a grid search over the minimum grain sizes $a_{\text{min}}$, the exponent of the power law that describes the grain size distribution $a_{\text{exp}}$ and grain composition. We assumed a grain size distribution of:
\begin{equation}
\frac{dN}{da} \sim \begin{cases}
     a^{-a_{\text{exp}}} & \text{if } a_{\text{min}} < a < a_{\text{max}}.\\
    0, & \text{otherwise}.
    \end{cases}
\end{equation}
We parameterized the grain composition in terms of the real and imaginary components indices of refraction of the dust grains. By doing so, we remain agnostic to the chemical composition of the grains. We also eliminate the need for the porosity parameter, whose effects are captured by the real and imaginary indices of the dust grain population assuming an uniform effective medium. 

\begin{table}
\begin{center}
\begin{tabular}{|l|c|c|c|l|}
\hline
Parameter & Start & End & Number of Points & Spacing\\
\hline
Minimum Grain Size ($\mu$m) & .01 & 100 & 15 & log \\
\hline
Grain Size Exponent & 2.5 & 6 & 10 & linear \\
\hline
Real Index of Refraction & 1.1 & 4.05 & 20 & linear \\
\hline
Imaginary Index of Refraction & $10^{-5}$ & 10. & 15 & log \\
\hline
Scattering law & \multicolumn{4}{|c|}{DHS/Mie} \\
\hline

\end{tabular}
\end{center}
\caption{Parameters for a grid search of different MCFOST models. 
The real and imaginary indices of refraction were chosen to reflect limits seen in physical grain models at the central wavelength of 2.15$\mu m$ \label{tab:grid_params}}
\end{table}

The ranges of our fitting parameters are shown in Table~\ref{tab:grid_params}. The phase curves were integrated over a range from $a_{\text{min}}$ to $a_{\text{max}}$. The maximum grain size, $a_{\text{max}}$ was fixed at 1\,mm due to the sharp power law which dictates that there are few large grains for any of the proposed grain size distribution. The limits of the real and imaginary indices of refraction were gained from the ranges of the indices for physical grain compositions at the \kb-band wavelength. Measured real and imaginary indices for a variety of different materials at \kb band are shown in Figure~\ref{fig:indices}. Whereas the usually assumed exponent $a_{\text{exp}}$ for the grain size power law distribution is usually assumed to be 3.5 following \cite{mathis97}, we fit over 10 different power laws. 

\subsection{Dust Grain Modelling Results\label{sec:dust_mod_results}}
We evaluated both DHS and Mie models for the grid defined in Table~\ref{tab:grid_params}. We examined the results of the resultant curves for each model with the metrics of the lowest $\chi_\nu$ for the total intensity curve, lowest $\chi_\nu$ for the polarized intensity curve, and lowest $\chi_\nu$ for both curves simultaneously. The ideal model needs the three distinctive properties of the HR4796A model: a strong forward scattering peak in the total intensity curve, a gradual increase in the total intensity curve at the backscattering side, and a peak in polarized intensity at 40$^\circ$.  

\begin{figure}
    \centering
    \includegraphics[width = 
    \textwidth]{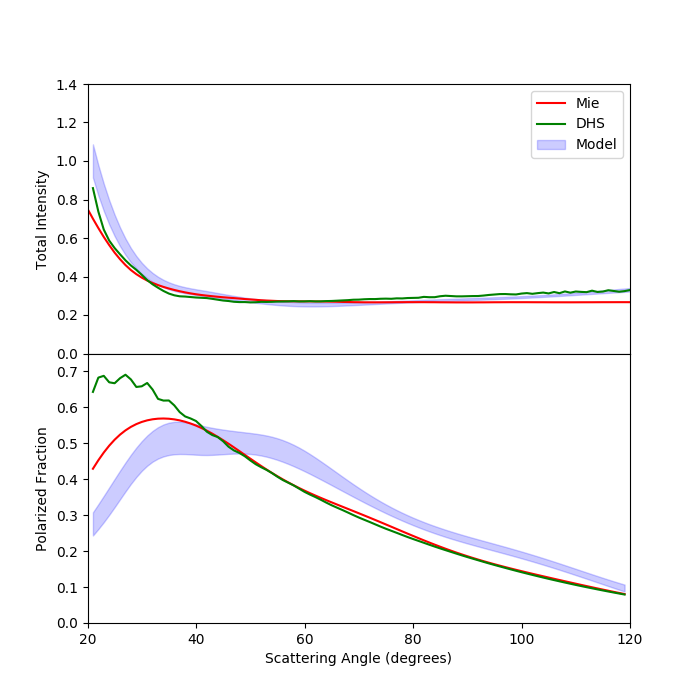}
    \caption{MCFOST fits of the \kb-band total intensity (upper) and polarized fraction (lower). The data-extracted curves and associated errors are shown in blue, while the red shows the best-fit curve for DHS and the green shows the best-fit curve for Mie. 
    \label{fig:phasefitscomb}}
\end{figure}

We found that neither model could simultaneously reproduce all of the features of both the SPF and polarized phase function. Figure~\ref{fig:phasefitscomb} shows the best-fit models for the simultaneous $\chi_\nu$. Though not a close model in total intensity, the DHS model is able to reproduce the features of the forward scattering peak as well as the shape of the curve in the backscattered direction. On the other hand, the DHS model is unable to reproduce the peak in polarized fraction at $40^\circ$. 

While the Mie model generates a polarized fraction curve with a peak closer to that of the data, the Mie model fails to recover the magnitude of the peak in total intensity as well as the increase in intensity at backscattering angles, exhibiting instead a flat distribution at scattering angles greater than $40^\circ$.

\begin{figure}
    \centering
    \includegraphics[width=0.8\textwidth]{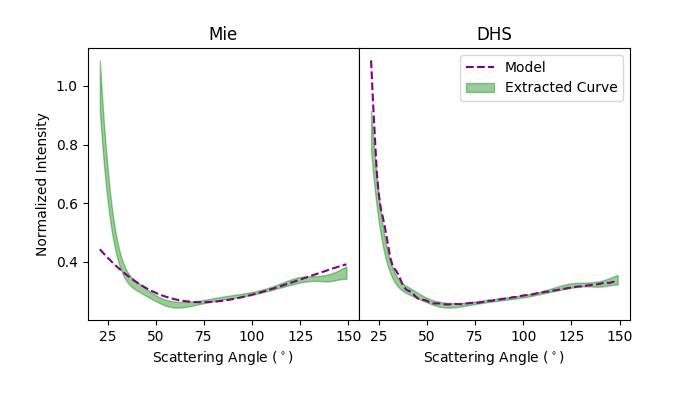}
    \caption{Best fit total intensity phase curves fitting only to \kb-band total intensity data for Mie theory (left) and DHS (right). }
    
    \label{fig:phasefitstot}
\end{figure}

We computed the goodness-of-fit metrics for the total intensity and polarized fraction phase curves independently of each other by calculating the $\chi_\nu^2$ of each ignoring the other. The models with the lowest $\chi_\nu$ of the total intensity phase curves are shown in Figure~\ref{fig:phasefitstot} with the parameters listed in Table~\ref{tab:best_phase}. In this case, the best-fit Mie model is able to reproduce the back scattering increase, but cannot produce a forward scattering peak sharp enough to match the model. The DHS model has a good fit to the overall curve, with a $\chi_\nu^2$ under 1. A comparison between the DHS $\chi^2_\nu$ of the total intensity-only fit and the $\chi_\nu^2$ of the combined polarized fraction and total intensity fit, shown in rows 5 and 4, respectively, show that the total intensity fit has an improvement on the total intensity $\chi^2$, but a drastically worse polarized fraction. 

\begin{figure}
    \centering
    \includegraphics[width=0.8\textwidth]{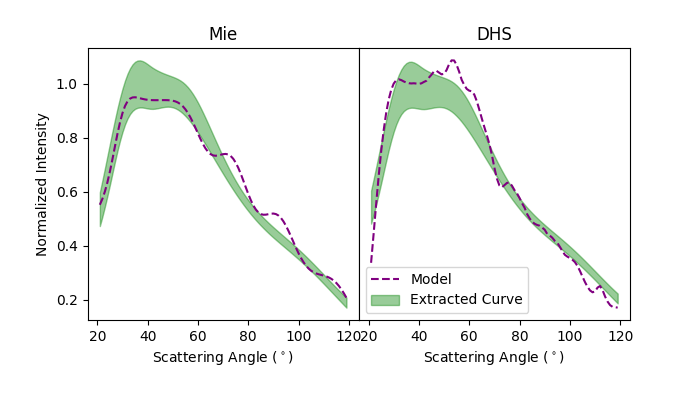}
    \caption{Best fit polarized fraction phase curves to only the \kb-band polarized fraction for Mie (left) and DHS (right).}
    \label{fig:phasefitsfrac}
\end{figure}

The curves produced by fitting only to the polarized fraction are shown in Figure~\ref{fig:phasefitsfrac}. Both curves have similar overall structures to the data phase curves with the peak at $40^\circ$, but exhibit unexpectedly an unexpectedly jagged curve. An examination of images produced by MCFOST with these phase curves do not visibly show any of this roughness given the pixel sampling and PSF convolution. Our model, constructed assuming a smooth phase curve would therefore be unable to detect any extra structure on the curve without overfitting.

\begin{table} 
\begin{center}
\begin{tabular}{|c|c|c|c|c|c|c|c|c|}
\hline
Grain Model & Metric $\chi_\nu^2$ & Summed $\chi_\nu^2$ & Tot Intensity $\chi_\nu^2$ & Pol Frac $\chi_\nu^2$ & $a_{\text{min}}$ & $a_{\text{exp}}$ & $n$ & $k$ \\
\hline
Mie & Sum & \textbf{12.4} & 13.3 & 11.8 & 1.9 & 2.9 & 3.7 & $3.72$ \\ 
\hline
Mie & Tot Intensity & 155.8 & \textbf{3.3} & 363.1 & 0.3 & 5.7 & 1.1 & 10.0 \\
\hline
Mie & Pol Frac & 165.2 & 293.6 & \textbf{3.0} & 3.7 & 3.7 & 2.0 & $2.7 \times 10^{-2}$ \\
\hline
DHS & Sum & \textbf{11.5} & 5.2 & 20.4 & 13.9 & 3.3 & 3.4 & $3.7$  \\
\hline
DHS & Tot Intensity & 41.2 & \textbf{0.8} & 96.2 & 26.8 & 4.1 & 1.1 & 1.4 \\
\hline
DHS & Pol Frac & 815.3 & 1458.8 & \textbf{3.0} & 1.9 & 2.9 & 3.4 & $3.0 \times 10^{-2}$ \\ 
\hline
\end{tabular}
\end{center}
\caption{Best fit parameters for different $\chi^2_\nu$ for different grain models. The bolded column indicates the $\chi^2_\nu$ each set of parameters was optimized for. The third column shows the sum of the total intensity and polarized fraction $\chi_\nu^2$s. The fourth column and fifth columns are the $\chi_\nu^2$ values for the total and polarized fraction, respectively. The second column describes the metric over which the best fit parameters were derived. The second and fifth rows list the best fit parameters in the sixth to ninth columns for the best summed $\chi_\nu^2$. The third and sixth columns are the parameters for the best total intensity $\chi_\nu^2$ and the fourth and sixth list the best fit parameters for the best polarized intensity.  \label{tab:best_phase}}
\end{table}

\subsection{Grain Indices of Refraction\label{sec:grain_discussion}}
\begin{figure}
\centering
        \includegraphics[width = 0.7\textwidth]{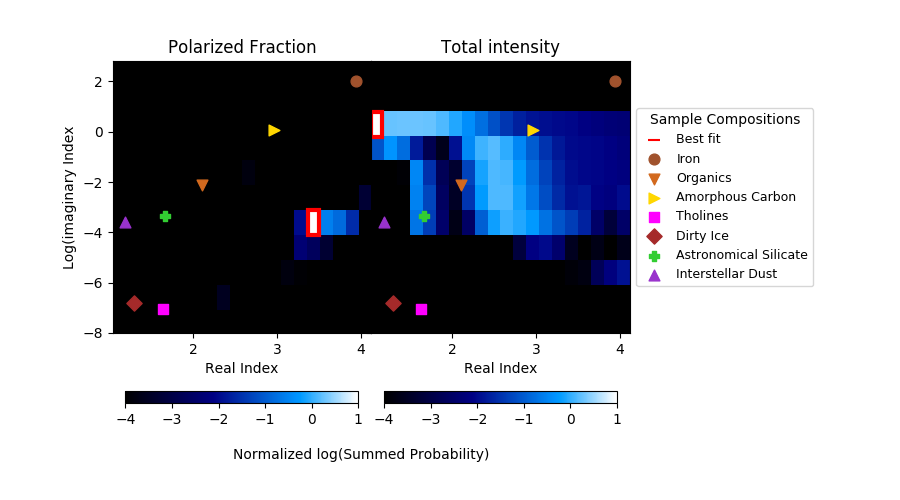}
    
    \caption{Marginalized probability maps of the DHS (upper) and (Mie) models for total intensity (left) and polarized fraction(right), gained by summing along the probability matrix along the $a_{\text{min}}$ and $a_{\text{exp}}$ axes. Overplotted are indices of refraction of representative dust grains 
    \citep{khare84,pollack94,zubko96,li97,li98,draineli2001}}
    \label{fig:indices}
\end{figure}

In order to further evaluate the generated DHS model compared to more physical models, we compared the phase space of likely indices of refraction we derived from our fits to indices of various other materials in Figure~\ref{fig:indices}. Following Bruggeman mixing rules, mixtures of two or more materials result in indices intermediate to the indices of the materials being mixed. A mixture of any dust compositions would lie somewhere along the semi-linear track traced out by the materials already shown. Porosity, essentially a mixture of void with a dust grain composition, would additionally move any point along the same track. 

The polarized fraction DHS best fits, boxed in red occupies a part of parameter space that is not only far from any pure dust grain composition, but would also be far from any mixture with any porosity. The parameter space of decent total intensity fits using DHS is fairly broad and overlaps with the track of physical compositions. However, the lack of overlap between the polarized fraction fits and the total intensity fits precludes any confident conclusions about the grain composition derived from the DHS fits.

\subsection{Discussion \label{sec:grain_disc}}
Both the Mie and DHS models are meant to be substitutes for more realistic, but more computationally expensive, models of aggregate dust grains. These aggregate dust grain models get exponentially more expensive with grain size. Our models that produce the smallest $\chi_\nu^2$ values all exhibit large grain sizes of 2 - 26 \micron, for which aggregate models have not been extensively generated. This analysis questions the validity of Mie and DHS models in producing meaningful results in this limit. The phase curves for HR4796A are unlike other phase curves in the defining features of the sharp total intensity forward scattering peak at 25$^\circ$, the modest backscattering peak and the polarized intensity peak at 25$^\circ$. Neither model was able to fully produce all three features simultaneously. 

There are a number of other ways that the dust population model can be improved in future work regarding the parameterization of the size distribution. Most obviously unphysical is the sharp cutoffs of our dust grain size distribution at the minimum and maximum grain sizes. Given the steep outer power law $a_{\text{exp}}$ and our large maximum grain size it is unlikely that increasing the maximum grain size cutoff would appreciably affect the resultant model. On the other hand, creating a more gradual distribution of grains rather than one that sharply cuts off at the minimum grain size would likely affect the model phase curves. It is likely that the jagged polarized fraction phase curves would be smoothed by the inclusion of smaller grains, but not without affecting the goodness of fit to the DHS total intensity model. Another major possibility is that there is not only a mixture of grains with different compositions, but also multiple dust grain populations with different size distributions. 

\section{CONCLUSIONS}\label{sec:conclusions}
We have presented high contrast polarimetry images of HR4796A in \kb- and \jb-band. Using a forward modeled disk to the polarized and total intensity, we have confirmed and put tighter constraints on the geometric properties of the disk. 

The unique features of the HR4796A disk and the high signal-to-noise of our data provides some of the tightest constraints on the properties of a dust grain population, where analyses of other disks result in degenerate solutions. These constraints have allowed us to completely eliminate a parameterization of Mie and DHS theories with any dust grain composition, and any porosity as an accurate model of the scattering dust grains. Further studies would necessarily need a model better model with more realistic features such as a more physical dust grain geometry, or a more complicated dust grain population.

For future studies, we additionally defer analysis of the chromaticity of the polarized phase function, which evidently extends from the visible to our measurements in the near-infrared. This analysis would necessitate a better extraction of the total intensity phase function of the \jb-band and the \hb-band.

\acknowledgements
This research was supported in part by NSF AST-1413718 and AST-1615272, AST-NNX15AC89G, AST-1413718, AST-1615272

J.M. acknowledges support for this work was provided by NASA through the NASA Hubble Fellowship grant \#HST-HF2-51414.001 awarded by the Space Telescope Science Institute, which is operated by the Association of Universities for Research in Astronomy, Inc., for NASA, under contract NAS5-26555.

Portions of this work were performed under the auspices of the U.S. Department of Energy by Lawrence Livermore National Laboratory under Contract DE-AC52-07NA27344.

This work is based on observations obtained at the Gemini Observatory, which is operated by the Association of Universities for Research in Astronomy, Inc. (AURA), under a cooperative agreement with the National Science Foundation (NSF) on behalf of the Gemini partnership: the NSF (United States), the National Research Council (Canada), CONICYT (Chile), Ministerio de Ciencia, Tecnolog\'ia e Innovaci\'on Productiva (Argentina), and Minist\'erio da Ci\^encia, Tecnologia e Inova\c c\~ao (Brazil). 

This work was supported by the NSF AST-1518332 (T.M.E., R.J.D.R., J.R.G., P.K., G.D.) and NASA grants NNX15AC89G and NNX15AD95G/NExSS (T.M.E., B.M., R.J.D.R., G.D., J.J.W, J.R.G., P.K.). This work benefited from NASA's Nexus for Exoplanet System Science (NExSS) research coordination network sponsored by NASA's Science Mission Directorate. 

GD acknowledge support from NSF grants AST-141378 and AST-1518332 and NASA grants NNX15AC89G and NNX15AD95G/NExSS.

\software{Gemini Planet Imager Data Pipeline \citep{perrin14}, PyKLIP~\citep{wang2014}, MCFOST~\citep{pinte06}, emcee~\citep{emcee}}
\bibliography{literature}

\newcommand{\noopsort}[1]{}
\begin{thebibliography}{}
\expandafter\ifx\csname natexlab\endcsname\relax\def\natexlab#1{#1}\fi

\bibitem[{{Bell} {et~al.}(2015){Bell}, {Mamajek}, \& {Naylor}}]{bell2015}
{Bell}, C.~P.~M., {Mamajek}, E.~E., \& {Naylor}, T. 2015, \mnras, 454, 593

\bibitem[{{Beuzit} {et~al.}(2019){Beuzit}, {Vigan}, {Mouillet}, {Dohlen},
  {Gratton}, {Boccaletti}, {Sauvage}, {Schmid}, {Langlois}, {Petit},
  {Baruffolo}, {Feldt}, {Milli}, {Wahhaj}, {Abe}, {Anselmi}, {Antichi},
  {Barette}, {Baudrand}, {Baudoz}, {Bazzon}, {Bernardi}, {Blanchard}, {Brast},
  {Bruno}, {Buey}, {Carbillet}, {Carle}, {Cascone}, {Chapron}, {Chauvin},
  {Charton}, {Claudi}, {Costille}, {De Caprio}, {Delboulb{\'e}}, {Desidera},
  {Dominik}, {Downing}, {Dupuis}, {Fabron}, {Fantinel}, {Farisato},
  {Feautrier}, {Fedrigo}, {Fusco}, {Gigan}, {Ginski}, {Girard}, {Giro},
  {Gisler}, {Gluck}, {Gry}, {Henning}, {Hubin}, {Hugot}, {Incorvaia}, {Jaquet},
  {Kasper}, {Lagadec}, {Lagrange}, {Le Coroller}, {Le Mignant}, {Le Ruyet},
  {Lessio}, {Lizon}, {Llored}, {Lundin}, {Madec}, {Magnard}, {Marteaud},
  {Martinez}, {Maurel}, {M{\'e}nard}, {Mesa}, {M{\"o}ller-Nilsson}, {Moulin},
  {Moutou}, {Orign{\'e}}, {Parisot}, {Pavlov}, {Perret}, {Pragt}, {Puget},
  {Rabou}, {Ramos}, {Reess}, {Rigal}, {Rochat}, {Roelfsema}, {Rousset}, {Roux},
  {Saisse}, {Salasnich}, {Santambrogio}, {Scuderi}, {Segransan}, {Sevin},
  {Siebenmorgen}, {Soenke}, {Stadler}, {Suarez}, {Tiph{\`e}ne}, {Turatto},
  {Udry}, {Vakili}, {Waters}, {Weber}, {Wildi}, {Zins}, \& {Zurlo}}]{beuzit19}
{Beuzit}, J.~L., {Vigan}, A., {Mouillet}, D., {et~al.} 2019, arXiv e-prints,
  arXiv:1902.04080

\bibitem[{{Bowler} \& {Nielsen}(2018)}]{bowler18}
{Bowler}, B.~P., \& {Nielsen}, E.~L. 2018, ArXiv e-prints, arXiv:1802.10132

\bibitem[{{Foreman-Mackey} {et~al.}(2013){Foreman-Mackey}, {Hogg}, {Lang}, \&
  {Goodman}}]{emcee}
{Foreman-Mackey}, D., {Hogg}, D.~W., {Lang}, D., \& {Goodman}, J. 2013, \pasp,
  125, 306

\bibitem[{{Gaia Collaboration} {et~al.}(2016){Gaia Collaboration}, {Prusti},
  {de Bruijne}, {Brown}, {Vallenari}, {Babusiaux}, {Bailer-Jones}, {Bastian},
  {Biermann}, {Evans}, \& et~al.}]{gaia16}
{Gaia Collaboration}, {Prusti}, T., {de Bruijne}, J.~H.~J., {et~al.} 2016,
  \aap, 595, A1

\bibitem[{{Greaves} {et~al.}(2000){Greaves}, {Mannings}, \&
  {Holland}}]{greaves00}
{Greaves}, J.~S., {Mannings}, V., \& {Holland}, W.~S. 2000, \icarus, 143, 155

\bibitem[{{Hinkley} {et~al.}(2009){Hinkley}, {Oppenheimer}, {Soummer},
  {Brenner}, {Graham}, {Perrin}, {Sivaramakrishnan}, {Lloyd}, {Roberts}, \&
  {Kuhn}}]{hinkley09}
{Hinkley}, S., {Oppenheimer}, B.~R., {Soummer}, R., {et~al.} 2009, \apj, 701,
  804

\bibitem[{{Jura}(1991)}]{jura1991}
{Jura}, M. 1991, \apjl, 383, L79

\bibitem[{{Khare} {et~al.}(1984){Khare}, {Sagan}, {Arakawa}, {Suits},
  {Callcott}, \& {Williams}}]{khare84}
{Khare}, B.~N., {Sagan}, C., {Arakawa}, E.~T., {et~al.} 1984, \icarus, 60, 127

\bibitem[{{Koerner} {et~al.}(1998){Koerner}, {Ressler}, {Werner}, \&
  {Backman}}]{koerner98}
{Koerner}, D.~W., {Ressler}, M.~E., {Werner}, M.~W., \& {Backman}, D.~E. 1998,
  \apjl, 503, L83

\bibitem[{{Konopacky} {et~al.}(2014){Konopacky}, {Thomas}, {Macintosh},
  {Dillon}, {Sadakuni}, {Maire}, {Fitzgerald}, {Hinkley}, {Kalas}, {Esposito},
  {Marois}, {Ingraham}, {Marchis}, {Perrin}, {Graham}, {Wang}, {De Rosa},
  {Morzinski}, {Pueyo}, {Chilcote}, {Larkin}, {Fabrycky}, {Goodsell},
  {Oppenheimer}, {Patience}, {Saddlemyer}, \&
  {Sivaramakrishnan}}]{konopacky2014}
{Konopacky}, Q.~M., {Thomas}, S.~J., {Macintosh}, B.~A., {et~al.} 2014, in
  Society of Photo-Optical Instrumentation Engineers (SPIE) Conference Series,
  Vol. 9147, Ground-based and Airborne Instrumentation for Astronomy V, 914784

\bibitem[{{Lagrange} {et~al.}(2010){Lagrange}, {Bonnefoy}, {Chauvin}, {Apai},
  {Ehrenreich}, {Boccaletti}, {Gratadour}, {Rouan}, {Mouillet}, {Lacour}, \&
  {Kasper}}]{lagrange10}
{Lagrange}, A.-M., {Bonnefoy}, M., {Chauvin}, G., {et~al.} 2010, Science, 329,
  57

\bibitem[{{Li} \& {Draine}(2001)}]{draineli2001}
{Li}, A., \& {Draine}, B.~T. 2001, \apj, 554, 778

\bibitem[{{Li} \& {Greenberg}(1997)}]{li97}
{Li}, A., \& {Greenberg}, J.~M. 1997, \aap, 323, 566

\bibitem[{{Li} \& {Greenberg}(1998)}]{li98}
---. 1998, \aap, 331, 291

\bibitem[{{Li} \& {Lunine}(2003)}]{li03}
{Li}, A., \& {Lunine}, J.~I. 2003, \apj, 590, 368

\bibitem[{{Lisse} {et~al.}(2017){Lisse}, {Sitko}, {Marengo}, {Vervack},
  {Fernandez}, {Mittal}, \& {Chen}}]{lisse17}
{Lisse}, C.~M., {Sitko}, M.~L., {Marengo}, M., {et~al.} 2017, \aj, 154, 182

\bibitem[{{Macintosh} {et~al.}(2006){Macintosh}, {Graham}, {Palmer}, {Doyon},
  {Gavel}, {Larkin}, {Oppenheimer}, {Saddlemyer}, {Wallace}, {Bauman}, {Evans},
  {Erikson}, {Morzinski}, {Phillion}, {Poyneer}, {Sivaramakrishnan}, {Soummer},
  {Thibault}, \& {Veran}}]{macintosh06}
{Macintosh}, B., {Graham}, J., {Palmer}, D., {et~al.} 2006, in Society of
  Photo-Optical Instrumentation Engineers (SPIE) Conference Series, Vol. 6272,
  62720L

\bibitem[{{Marois} {et~al.}(2006){Marois}, {Lafreni{\`e}re}, {Doyon},
  {Macintosh}, \& {Nadeau}}]{marois06}
{Marois}, C., {Lafreni{\`e}re}, D., {Doyon}, R., {Macintosh}, B., \& {Nadeau},
  D. 2006, \apj, 641, 556

\bibitem[{{Mathis} {et~al.}(1977){Mathis}, {Rumpl}, \& {Nordsieck}}]{mathis97}
{Mathis}, J.~S., {Rumpl}, W., \& {Nordsieck}, K.~H. 1977, \apj, 217, 425

\bibitem[{{Mayor} \& {Queloz}(1995)}]{mayor95}
{Mayor}, M., \& {Queloz}, D. 1995, \nat, 378, 355

\bibitem[{{Milli} {et~al.}(2017){Milli}, {Vigan}, {Mouillet}, {Lagrange},
  {Augereau}, {Pinte}, {Mawet}, {Schmid}, {Boccaletti}, {Matr{\`a}}, {Kral},
  {Ertel}, {Chauvin}, {Bazzon}, {M{\'e}nard}, {Beuzit}, {Thalmann}, {Dominik},
  {Feldt}, {Henning}, {Min}, {Girard}, {Galicher}, {Bonnefoy}, {Fusco}, {de
  Boer}, {Janson}, {Maire}, {Mesa}, {Schlieder}, \& {SPHERE
  Consortium}}]{milli2017}
{Milli}, J., {Vigan}, A., {Mouillet}, D., {et~al.} 2017, \aap, 599, A108

\bibitem[{{Milli} {et~al.}(2019){Milli}, {Engler}, {Schmid}, {Olofsson},
  {Menard}, {Kral}, {Boccaletti}, {Thebault}, {Choquet}, {Mouillet},
  {Lagrange}, {Augereau}, {Pinte}, {Chauvin}, {Dominik}, {Perrot}, {Zurlo},
  {Henning}, {Min}, {Beuzit}, {Avenhaus}, {Bazzon}, {Moulin}, {Llored},
  {Moeller-Nilsson}, {Roelfsema}, \& {Pragt}}]{milli19}
{Milli}, J., {Engler}, N., {Schmid}, H.~M., {et~al.} 2019, arXiv e-prints,
  arXiv:1905.03603

\bibitem[{{Min} {et~al.}(2005){Min}, {Hovenier}, \& {de Koter}}]{min2005}
{Min}, M., {Hovenier}, J.~W., \& {de Koter}, A. 2005, \aap, 432, 909

\bibitem[{{Nesvold} \& {Kuchner}(2015)}]{nesvold15}
{Nesvold}, E.~R., \& {Kuchner}, M.~J. 2015, \apj, 798, 83

\bibitem[{{Olofsson} {et~al.}(2019){Olofsson}, {Milli}, {Th{\'e}bault}, {Kral},
  {M{\'e}nard}, {Janson}, {Augereau}, {Bayo}, {Beam{\'\i}n}, {Henning},
  {Iglesias}, {Kennedy}, {Montesinos}, {Pawellek}, {Schreiber}, {Zamora},
  {Carbillet}, {Feautrier}, {Fusco}, {Madec}, {Rabou}, {Sevin}, {Szul{\'a}gyi},
  \& {Zurlo}}]{olofsson2019}
{Olofsson}, J., {Milli}, J., {Th{\'e}bault}, P., {et~al.} 2019, \aap, 630, A142

\bibitem[{{Perrin} {et~al.}(2014){Perrin}, {Maire}, {Ingraham}, {Savransky},
  {Millar-Blanchaer}, {Wolff}, {Ruffio}, {Wang}, {Draper}, {Sadakuni},
  {Marois}, {Rajan}, {Fitzgerald}, {Macintosh}, {Graham}, {Doyon}, {Larkin},
  {Chilcote}, {Goodsell}, {Palmer}, {Labrie}, {Beaulieu}, {De Rosa},
  {Greenbaum}, {Hartung}, {Hibon}, {Konopacky}, {Lafreniere}, {Lavigne},
  {Marchis}, {Patience}, {Pueyo}, {Rantakyr{\"o}}, {Soummer},
  {Sivaramakrishnan}, {Thomas}, {Ward-Duong}, \& {Wiktorowicz}}]{perrin14}
{Perrin}, M.~D., {Maire}, J., {Ingraham}, P., {et~al.} 2014, in \procspie, Vol.
  9147, Ground-based and Airborne Instrumentation for Astronomy V, 91473J

\bibitem[{{Perrin} {et~al.}(2015){Perrin}, {Duchene}, {Millar-Blanchaer},
  {Fitzgerald}, {Graham}, {Wiktorowicz}, {Kalas}, {Macintosh}, {Bauman},
  {Cardwell}, {Chilcote}, {De Rosa}, {Dillon}, {Doyon}, {Dunn}, {Erikson},
  {Gavel}, {Goodsell}, {Hartung}, {Hibon}, {Ingraham}, {Kerley}, {Konapacky},
  {Larkin}, {Maire}, {Marchis}, {Marois}, {Mittal}, {Morzinski}, {Oppenheimer},
  {Palmer}, {Patience}, {Poyneer}, {Pueyo}, {Rantakyr{\"o}}, {Sadakuni},
  {Saddlemyer}, {Savransky}, {Soummer}, {Sivaramakrishnan}, {Song}, {Thomas},
  {Wallace}, {Wang}, \& {Wolff}}]{perrin15}
{Perrin}, M.~D., {Duchene}, G., {Millar-Blanchaer}, M., {et~al.} 2015, \apj,
  799, 182

\bibitem[{{Pinte} {et~al.}(2006{\natexlab{a}}){Pinte}, {M{\'e}nard},
  {Duch{\^e}ne}, \& {Bastien}}]{pinte06}
{Pinte}, C., {M{\'e}nard}, F., {Duch{\^e}ne}, G., \& {Bastien}, P.
  2006{\natexlab{a}}, \aap, 459, 797

\bibitem[{{Pinte} {et~al.}(2006{\natexlab{b}}){Pinte}, {M{\'e}nard},
  {Duch{\^e}ne}, \& {Bastien}}]{mcfost}
---. 2006{\natexlab{b}}, \aap, 459, 797

\bibitem[{{Pollack} {et~al.}(1994){Pollack}, {Hollenbach}, {Beckwith},
  {Simonelli}, {Roush}, \& {Fong}}]{pollack94}
{Pollack}, J.~B., {Hollenbach}, D., {Beckwith}, S., {et~al.} 1994, \apj, 421,
  615

\bibitem[{{Pueyo}(2016)}]{pueyo16}
{Pueyo}, L. 2016, \apj, 824, 117

\bibitem[{{Quillen}(2006{\natexlab{a}})}]{quillen06}
{Quillen}, A.~C. 2006{\natexlab{a}}, \mnras, 372, L14

\bibitem[{{Quillen}(2006{\natexlab{b}})}]{kalas05}
---. 2006{\natexlab{b}}, \mnras, 372, L14

\bibitem[{{Rodigas} {et~al.}(2015){Rodigas}, {Stark}, {Weinberger}, {Debes},
  {Hinz}, {Close}, {Chen}, {Smith}, {Males}, {Skemer}, {Puglisi}, {Follette},
  {Morzinski}, {Wu}, {Briguglio}, {Esposito}, {Pinna}, {Riccardi}, {Schneider},
  \& {Xompero}}]{rodigas15}
{Rodigas}, T.~J., {Stark}, C.~C., {Weinberger}, A., {et~al.} 2015, \apj, 798,
  96

\bibitem[{{Schneider} {et~al.}(2009){Schneider}, {Weinberger}, {Becklin},
  {Debes}, \& {Smith}}]{schneider2009}
{Schneider}, G., {Weinberger}, A.~J., {Becklin}, E.~E., {Debes}, J.~H., \&
  {Smith}, B.~A. 2009, \aj, 137, 53

\bibitem[{{Schneider} {et~al.}(1999){Schneider}, {Smith}, {Becklin}, {Koerner},
  {Meier}, {Hines}, {Lowrance}, {Terrile}, {Thompson}, \&
  {Rieke}}]{schneider1999}
{Schneider}, G., {Smith}, B.~A., {Becklin}, E.~E., {et~al.} 1999, \apjl, 513,
  L127

\bibitem[{{Schneider} {et~al.}(2014){Schneider}, {Grady}, {Hines}, {Stark},
  {Debes}, {Carson}, {Kuchner}, {Perrin}, {Weinberger}, {Wisniewski},
  {Silverstone}, {Jang-Condell}, {Henning}, {Woodgate}, {Serabyn},
  {Moro-Martin}, {Tamura}, {Hinz}, \& {Rodigas}}]{schneider14}
{Schneider}, G., {Grady}, C.~A., {Hines}, D.~C., {et~al.} 2014, \aj, 148, 59

\bibitem[{{Sheret} {et~al.}(2004){Sheret}, {Dent}, \& {Wyatt}}]{sheret04}
{Sheret}, I., {Dent}, W.~R.~F., \& {Wyatt}, M.~C. 2004, \mnras, 348, 1282

\bibitem[{{Soummer} {et~al.}(2012){Soummer}, {Pueyo}, \& {Larkin}}]{soummer12}
{Soummer}, R., {Pueyo}, L., \& {Larkin}, J. 2012, \apjl, 755, L28

\bibitem[{{van Leeuwen}(2007)}]{van_leeuwen2007}
{van Leeuwen}, F. 2007, \aap, 474, 653

\bibitem[{{Wang} {et~al.}(2015){Wang}, {Ruffio}, {De Rosa}, {Aguilar}, {Wolff},
  \& {Pueyo}}]{wang2015}
{Wang}, J.~J., {Ruffio}, J.-B., {De Rosa}, R.~J., {et~al.} 2015, {pyKLIP: PSF
  Subtraction for Exoplanets and Disks}, Astrophysics Source Code Library,
  ascl:1506.001

\bibitem[{{Wang} {et~al.}(2014){Wang}, {Rajan}, {Graham}, {Savransky},
  {Ingraham}, {Ward-Duong}, {Patience}, {De Rosa}, {Bulger},
  {Sivaramakrishnan}, {Perrin}, {Thomas}, {Sadakuni}, {Greenbaum}, {Pueyo},
  {Marois}, {Oppenheimer}, {Kalas}, {Cardwell}, {Goodsell}, {Hibon}, \&
  {Rantakyr{\"o}}}]{wang2014}
{Wang}, J.~J., {Rajan}, A., {Graham}, J.~R., {et~al.} 2014, in \procspie, Vol.
  9147, Ground-based and Airborne Instrumentation for Astronomy V, 914755

\bibitem[{{Wyatt}(2008)}]{wyatt08}
{Wyatt}, M.~C. 2008, \araa, 46, 339

\bibitem[{{Wyatt} {et~al.}(1999){Wyatt}, {Dermott}, {Telesco}, {Fisher},
  {Grogan}, {Holmes}, \& {Pi{\~n}a}}]{wyatt99}
{Wyatt}, M.~C., {Dermott}, S.~F., {Telesco}, C.~M., {et~al.} 1999, \apj, 527,
  918

\bibitem[{{Zubko} {et~al.}(1996){Zubko}, {Mennella}, {Colangeli}, \&
  {Bussoletti}}]{zubko96}
{Zubko}, V.~G., {Mennella}, V., {Colangeli}, L., \& {Bussoletti}, E. 1996,
  \mnras, 282, 1321

\end{thebibliography}

%
%

%
%



\end{document}